**Dynamic optical contrast imaging for real-time delineation of tumor resection margins using head and neck cancer as a model**

Yong Hu, PhD[1,2*]; Shan Huang, BS[3]; Albert Y. Han, MD, PhD[1,4,5]; Seong Moon, BS[1]; Jeffrey F. Krane, PhD[6]; Oscar Stafsudd, PhD[7]; Warren Grundfest[2]; Maie A. St. John, MD, PhD[1,4,5]

[1] Department of Head and Neck Surgery,
University of California, Los Angeles (UCLA), Los Angeles, California, USA
[2] Department of Bioengineering, UCLA, Los Angeles, CA, USA
[3] Department of Materials Science and Engineering, UCLA, Los Angeles, CA, USA
[4] UCLA Head and Neck Cancer Program, UCLA Medical Center, Los Angeles, CA, USA
[5] Jonsson Comprehensive Cancer Center, UCLA Medical Center, Los Angeles, CA, USA
[6] Department of Pathology & Laboratory Medicine, UCLA Medical Center, Los Angeles, CA, USA
[7] Department of Electrical and Computer Engineering, UCLA, Los Angeles, CA, USA

* This work was done by Yong Hu and co-authors when Yong was a graduate student and then a postdoc at UCLA. Yong finalized the manuscript at NUS, Singapore. Detailed author contributions could be referred to in the author contribution section.

*Corresponding to: yonghu@nus.edu.sg

**Abstract**

Complete surgical resection of the tumor for Head and neck squamous cell carcinoma (HNSCC) remains challenging, given the devastating side effects of aggressive surgery and the anatomic proximity to vital structures. To address the clinical challenges, we introduce a wide-field, label-free imaging tool that can assist surgeons delineate tumor margins real-time. We assume that autofluorescence lifetime is a natural indicator of the health level of tissues, and ratio-metric measurement of the emission-decay state to the emission-peak state of excited fluorophores will enable rapid lifetime mapping of tissues. Here, we describe the principle, instrumentation, characterization of the imager and the intraoperative imaging of resected tissues from 13 patients undergoing head and neck cancer resection. 20 x 20 $mm^2$ imaging takes 2 second/frame with a working distance of 50 mm, and characterization shows that the spatial resolution reached 70 μm and the least distinguishable fluorescence lifetime difference is 0.14 ns. Tissue imaging and Hematoxylin-Eosin stain slides comparison reveals its capability of delineating cancerous boundaries with submillimeter accuracy and a sensitivity of 91.86% and specificity of 84.38%.

**Main**

More than 1.5 million new cases of head and neck cancers were diagnosed, causing more than 510,000 deaths in 2020, worldwide.[1] Most of the cancers are debilitating diseases where patient prognosis depends heavily on complete and accurate tumor resection. The goal of cancer surgery is to remove the cancer in its entirety while minimizing the volume of resected healthy tissue to support patient function. The severe functional consequences of aggressive surgery may include loss of voice, as well as swallowing and speech problems. Despite new developments in surgical tools and techniques, persistent or recurrent disease after treatment develops in 65% of patients [2]. Currently, Magnetic Resonance Imaging (MRI), Computerized Tomography (CT) scans and other preoperative imaging techniques are used to assess tumor location, size, and extent (Figure 1a), but intraoperative margin detection relies solely on visual and tactile feedback during surgical resection. Final assessment of margins is performed via frozen section microscopy. The decision to terminate the resection or extend margins is then based on the combination of the preoperative imaging studies, the visual and tactile observations of the surgeon, and the results of frozen section pathology obtained after tumor resection (Figure 1b). The frozen section procedure is a pathological laboratory method to perform rapid microscopic analysis of a specimen. It is used most often in cancer surgery. After a tumor is presumed to be removed in its entirety, the surrounding tissues are sampled by frozen section to ensure that there is no microscopic disease left behind. However, this adds to duration of surgery, cost, and additional tissue processing. Furthermore, efficacy of the current approach varies widely and is subject to surgeon and pathologist's experience and sampling error.[3]

Prior efforts to use a variety of optical and ultrasonic techniques for margin detection have not succeeded clinically for many reasons. Optical technologies including optical coherence tomography and Raman spectroscopy have been investigated but system complexity, although a benefit in the laboratory in in vitro samples and phantoms, results in precipitous sensitivity to in vivo confounders, which render them suboptimal in an intraoperative setting.[4,5] Near-infrared imaging has demonstrated promising results but require the infusion of exogenous fluorophores.[6,7] Narrow band imaging has been widely available but hasn't gained trust from surgeons because of its relatively low diagnostic accuracy and long learning curve.[8-10] Intraoperative ultrasound for head and neck cancers have not yet shown a strong correlation between histologic and ultrasound estimates of depth of invasion.[11-13] Other optical techniques such as autofluorescence imaging (AFI), targeted fluorescence imaging (TFI), high-resolution microendoscopy (HRME), elastic scattering spectroscopy, confocal laser endomicroscopy, and confocal reflectance microscopy has been widely applied and reviewed.[14-18]Amongst all the studies, researchers have demonstrated that fluorescence lifetime imaging microscopy (FLIM) imaging has the potential either during surgery or on excised specimens to characterize biochemical features and associated pathologies.[19-21] FLIM has great advantages over quantitative imaging of intensities because lifetime values are less prone to intensity artefacts that are associated with changes in local concentration of fluorophores, local illumination intensity, target geometry, optical path, and detection efficiency.[22-24] However, despite the obvious advantages, FLIM met its own translational problems when groups tried to apply it to clinical applications. FLIM is usually implemented in a point scanning fashion and requires significant amount of time for both image capture and lifetime calculation.[25-29] As a result, the size of the field of view (FOV) and the spatial resolution are always a trade-off with image acquisition time. The limited depth of field, extremely short working distance and relatively narrow FOV are factors hindering the translational progress as well, because surgeons need an instrument that can freely navigate through the irregular surfaces of human tissues/organs. On the other hand, gated intensified charge coupled device (iCCD) based FLIMs have been previously reported but usually not quantitatively calibrated and generally suffered from poor accuracy or long imaging time.[30-34]

An intraoperative instrument that can significantly improve the accuracy of margin detection over current methods will improve outcomes for cancer patients by minimizing removal of normal functional tissue while ensuring complete tumor removal. Dynamic Optical Contrast Imaging (DOCI), system diagram shown in Figure 1c, uses a gated iCCD to compare the accumulated intensities during the emission-decay state to the emission-peak state, thus generating a spatially resolved map of relative differences in autofluorescence lifetime of tissues (Figure 1d). This approach ensures very fast processing since only three images are acquired and no deconvolution or exponential fitting required. The DOCI system features rapid, widefield, label-free imaging under room light environment with the use of light-emitting diodes (LEDs). This contrasts with previously explored FLIM that employs raster scanning, which results in long scanning time, small depth of field, and close working distance, which is impractical for intraoperative use. DOCI offers a new optical method characterized by speed, low cost, and improved accuracy, for enhancing intraoperative imaging and margin detection in all cancer patients. This intraoperative instrument is the first of its kind and has the potential to significantly improve the sensitivity and specificity of determining true head and neck cancer margins, enabling the surgeon to preserve healthy tissue and improve patient outcomes.

## Results

**Development of the imaging system.** The overall scheme of the experimental setup is summarized in Figure 1c. A LED, operating at a center wavelength of 368 nm, is driven by a custom-made circuit board that well regulates the emission intensity of the LED following the signal profile from the function generator (Keysight Agilent 33220A). The LED light is guided to a collimating lens using a Liquid Light Guide (LLG, Thorlabs), which provides excellent flexibility and better transmission at the interested spectrum range compared to traditional silica fiber bundles. The illumination light shines on the sample with a 45-degree incident angle to minimize possible reflectance to the camera. The fluorescent emission is collected using a Nikon compact-macro lens with a focal length of 50 mm that is coupled to a 10-slot motorized filter wheel (X-FWR06A-E02, Zaber). The 10 windows, a blank window, one long-pass filter (405+ nm), and followed by 8 bandpass filters centered at 415, 434, 465, 494, 520, 542, 572 and 605 nm, are switched for data collection in different spectral range. Each bandpass filter has an approximate 10 to 30 nm bandwidth and the filter set covers interested spectrum from 400 to 620 nm, as shown in Figure 2a. The filters will likely be changed, and number reduced for certain applications. An iCCD camera, Andor iStar 334T, is used for image acquisition and provides averaging in hardware.

The computer communicates with the function generator, filter wheel, and iCCD camera and is used for automation control, data processing, image display, as well as performance optimization for different scenarios. A photograph of the system with the illumination and imaging head inset is displayed in Figure 2b. Figure 2c shows the simulation results using ASAP (Breault Research Organization) of the intensity profile of the illumination light out of the LLG, which displays a homogenous circular and agrees well with the real-world illumination as shown in Figure 2d. The light out of the LLG is further expanded and collimated to better adjust to different working distances. Figure 2e displays the outlook of our customized graphic user interface (GUI). The GUI is designed to make the operation, data acquisition and image display as easy as possible. In the "Video" mode, video is acquired through the blank window to enable the operator locating the region of interest. The "Imaging" mode will automate the whole sequence of image acquisition through 9 windows of different filters, display lifetime images on the panel on the right and store 9 sets of reference/decay/background data within a timestamped file folder. The operator can also choose to operate the system in a "Manual" mode to allow direct control over all functions, such as observing lifetime video in the most interested window.

**Characterization of spatial resolution and temporal resolution.** Spatial resolution refers to the ability of an imaging modality to differentiate two adjacent structures as being distinct from one another and here it defines the smallest cluster of tumor cells that can be detected from the surrounding benign tissues. It is heavily dependent on the optics, noise level of the system, and calibration parameters. The calibrating FOV was set as 20 mm by 20 mm with a working distance of 50 mm. The calibration target used was a UV Fused Silica and Fluorescent USAF 1951 Resolution Target (Edmund Optics Inc.), which has a 365 nm excitation wavelength and a 550 nm emission wavelength. The calibration was conducted with room light on. Fluorescence intensity images, one as shown in Figure 3a, were used to generate the DOCI image in Figure 3b. Even though the resolution target is coated with the same fluorescent material, the intensity map varies because of the relative illumination intensity changes. However, the DOCI image produced a uniform DOCI value (yellow) because the DOCI methodology ignores fluorescent intensity and reports only the fluorescence lifetime of the coating material. Some blue areas at the corners are treated as unreliable DOCI values by the algorithm, which is caused by insufficient illumination. The most closely spaced line pairs distinguishable on the DOCI image are 70 μm apart (Element 6 of Group 2); from this, we can infer that the spatial resolution of the system is 70 μm. In sum, the system can accurately detect the autofluorescence of abnormal cell clusters as small as 70 μm in diameter with a working distance of 50 mm and a FOV of 20 mm x 20 mm.

Temporal resolution refers to the precision of a measurement with respect to time and here it is defined as the smallest detectable lifetime variance that is unique to our system. Three types of dyes at two concentrations were used to validate the DOCI system. The three types of dye solutions were Laurdan dissolved in dimethyl sulfoxide (DMSO), NADH in water, and 7-hydroxy-4-methylcoumarin in methanol. Each dye solution was prepared in two different concentrations. A drop from each of the six solutions was placed onto a microscope slide, and they were imaged under the same field of view by the DOCI system (Figure 3c). The fluorescence intensity images were captured during the decaying period and the reference period as shown in the top and bottom figures in Figure 3d. Brightness variances amongst drops and across the periphery of each drop are visible, which are mainly due to different dye concentrations, various fluorescence lifetimes, geometric variances, and inhomogeneous illumination.

The corresponding relative lifetime image in Figure 3e was captured with DOCI through window 9 (597-613 nm). The DOCI image shows only three colors for the three types of dye solutions, regardless of the changes in dye concentrations, drop geometries, and illumination intensities. The specific 50-by-50-pixel regions of interest were selected on each drop to calculate the aggregate means and standard deviations of the relative lifetimes measured (Figure 3g). Nine filters designed for different wavelength ranges also exhibited identical DOCI values (Figure 3f), supporting the claim that the DOCI system can extract fluorescence lifetime measurements independent from dye concentrations, geometrical variances, and illumination intensities. This is critical for clinical translation, in which a robust fluorescence lifetime imaging system had better yield images only based on fluorescence lifetimes despite uneven fluorophore distributions, irregular surfaces and inhomogeneous illuminations which are often encountered in the clinical and surgical settings.

The mathematical model for DOCI (see Methods) illustrates a linear relationship between DOCI values and autofluorescence lifetimes with suitable gating width. We experimentally validated this model using a multiphoton confocal microscope (Leica Deep In Vivo Explorer SP8 DIVE). We prepared five types of solutions (see details in solution preparation in Methods) and each solution was split into two equal parts for simultaneous fluorescence lifetime measurement using multiphoton microscopy and DOCI. The scatter plot of the fluorescence lifetimes measured by FLIM and DOCI shows a linear correlation between the two variables as shown in Figure 3h. The linear fitting of the relative fluorescence lifetime from DOCI value (with a gating width of 20 ns) to FLIM results in a ratio (FLIM/DOCI) of 21.03 (1/Slope, Figure 3h). As we had an average standard deviation of 0.0068 from the sampled measurements in Figure 3g, the temporal resolution of the system is 0.14 ns, calculated by multiplying the average standard deviation by the ratio. Linear relationship between DOCI values and absolute lifetime values can be achieved with proper parameter settings (see details in mathematical model), which could turn DOCI into an absolute lifetime imager with further calibration.

**Intraoperative imaging of freshly sectioned diseased tissues from patients with head and neck squamous cell carcinoma (HNSCC).** 13 patients undergoing head and neck oncologic resection with curative intent were prospectively enrolled and they all received preoperative imaging examinations to get information such as size, location, and extent of the tumor. After surgical resection, specimens were immediately transferred by courier to UCLA Translational Pathology Core Laboratory (TPCL) for further tissue processing. The elapsed time between excision and arrival at the TPCL ranged from 10 to 30 minutes with a mean of approximately 25 minutes. We first take an image of the surface of the specimen using DOCI and then the specimens were sectioned into multiple fresh samples containing tumor and contiguous normal tissue of suspect lesions by designated staff at TPCL. The fresh samples are imaged with DOCI immediately and the image acquisition through all 9 windows requires approximately 15 seconds. After the sample was returned to the pathologist, the tissue was snap-frozen and subsequently sectioned for H&E stain by the pathology laboratory. Bright field images of the specimen were acquired with a digital single-lens reflex camera (Canon DSLR EOS Rebel T5i). After the completion of imaging, specimens were fixed in 10% buffered formalin and processed to paraffin. Blocks were oriented with reference to the macroscopic images and DOCI maps (fiducial) to correlate imaging and histopathology. A simplified protocol of tissue processing at TPCL and DOCI imaging is illustrated in Figure 4a. The head and neck surgeon were notified when the slides of the imaged specimen and the pathology report were available, and a multidisciplinary effort was made to demarcate the tissue type with visual cues from the visible images and histologic cues from the slides. This data labeling was used to drive the statistical analysis of contrast in the DOCI images.

To vividly demonstrate how a DOCI map could directly reflect the differences between cancerous and benign tissues, a representative case of squamous cell carcinoma (SCC) of the scalp was selected and processed in Figure 4. The white light image, fluorescence intensity image and corresponding DOCI image of the cross section were obtained (Figure 4b-d). Areas of carcinoma, fibrous tissue and normal squamous epithelium were denoted by a senior head and neck pathologist using serial slides stained with hematoxylin and eosin (H&E) (Figure 4e). The whole specimen was imaged with DOCI (Figure 4d), and representative areas of cancerous and non-cancerous fibrous tissues were obtained (Figures 4g). Three 50x50-pixel ROIs in the tumor area and four ROIs in the non-tumor areas were randomly selected (Figure 4.d). The mean and standard deviation of the DOCI values from all nine windows were plotted in Figure 4f. The scatter plot showed that the DOCI values of tumors at wavelengths of 415, 520, 542, 572 and 605 nm showed distinct values unlike those of normal surrounding tissues. This likely reflects the emission spectrum of tumor intrinsic fluorophores. In this case, the white light image of the specimen or autofluorescence intensity presented no visually distinguishable correlation to the histological ground truth, while the DOCI heat map (Figure 4d) reasonably reflected the triangle-like area of cancer that was annotated in Figure 4e. Even though DOCI value represents significant differences between cancer/benign tissues in 5 out of 10 wavelengths ($P<0.01$, t-test) and provides informal tissue maps, it still relies on the surgeon to make the final judgement. We continue to explore the auto segmentation in the following section.

**Cancer prediction based on multiple-channel lifetime information.** We designed a prospective clinical trial to test the validity of the DOCI technique for predicting cancer margins using multi-channel lifetime information and compared them to histology as the gold standard. Thirteen patients were enrolled in this prospective trial and 7 of them were used for analysis, excluding those samples contaminated by surgical ink, or had only cancerous tissues or benign tissues. The tissue processing and imaging followed the same protocol as described in the previous section. Upon collection of data, we used the basic linear discriminant classifier (MATLAB 2020a) to predict cancer using DOCI information from combinations of available channels and confined our model to two-class classification and evaluation.

One representative case of a patient's SCC ear is selected to illustrate the prediction process and outcome in Figure 5. The specimens (Figure 5a) were imaged with the DOCI technique, and the specimen was subsequently sectioned parallel to the imaged plane for H&E staining as the ground truth. A senior head and neck pathologist annotated the H&E slide in Figure 5b. In Figure 5d, the fluorescence intensity image was annotated manually based on the H&E slides and a series of small random ROIs were chosen in different areas (pink for background corkboard, purple for cartilage, yellow for healthy fibrous tissue, and green for cancer) for classifier training. The DOCI prediction map for this case is displayed in Figure 5e and it is resampled to 0.65mm x 0.65mm blocks, whose size makes more sense in surgical scenario in the OR than single pixels, for efficacy evaluation using the H&E as gold standard. The predicted tissue types were color-coded, where cyan delineates the boundary of the tissue in the FOV, blue as annotated cancerous area based on the H&E results, red as predicted cancerous area, and purple is the overlap of blue and red, which stands for correct cancer prediction. Figures 5c and 5f are zoomed-in areas that are red-boxed in Figures 5b and 5e, showing details of the H&E results and cancer prediction of the same part of the specimen. Similar raw and processed images of three more representative cases from different sites of diseases are available in Extended Figure 1.

Each tissue prediction made using the linear discriminant classifier was compared with the H&E histological ground truth. In the comparison, if any annotated cancer falls in a block, the block is treated as true cancer and those without any annotated cancer as true benign; similarly, if any predicted cancer falls in a block, the block is considered predicted cancer and those without predicted cancer as predicted benign. Consequently, if a block is true benign and predicted benign at the same time, it is true negative (TN), colored as cyan; if a block is true cancer but predicted benign, then it is false negative (FN), colored as blue; if a block is true cancer and predicted cancer at the same time, it is true positive (TP), colored as purple; lastly, if a block is true benign but predicted cancer, it is false positive (FP), colored as red. Data from each of the 9 channels were used for classifier training and the efficacy of predication is evaluated subsequently for each channel. The same process was applied to expanded data of two channels, which has 36 ($C_9^2$) combinations, data of three channels that has 84 ($C_9^3$) combinations, and all data from 9 channels. This analysis included a total aggregate image surface area of 1559 $mm^2$ from seven patients out of the 13 enrolled, consisting of 3691 blocks. The accuracy of the predictions, defined as the number of TN and TP blocks divided by the total number of blocks, with the highest scores from nine channels, three channels and two channels were 88.81%, 88.00% and 86.56%, respectively, as shown in Figure 5g. A more complete efficacy statistics could be seen in extended data Table

1. The results of the prospective trial showed that the DOCI system provides a greater precision compared to visual or tactile feedback. Furthermore, the systematic nature of the analysis pipeline presents a rapid, reliable alternative to frozen section analysis.

**Discussion**

Current clinically adopted methods of tumor tissue localization are limited to visual inspection and tactile assessment. The success of these methods is strongly operator-dependent and affected mostly by training and other human factors that are difficult to standardize. Furthermore, frozen section analysis informed by visual and tactile assessment is subject to significant sampling errors and results in extensive human and financial costs arising from frequent false positives and false negatives.[35] The majority of diagnostic system architectures under research for intraoperative tumor localization are optical imaging-based systems, as this is the preferred approach for circumventing sampling error. However, most of these systems require exogenous fluorophores in the form of injectable contrast agents and have failed to translate clinically due to a variety of practical constraints and performance deficiencies. The DOCI technique is a result of this thinking; it is rapid, only uses endogenous fluorophores, and is robust to confounders common to clinical and intraoperative settings.

The key parameter utilized by DOCI technique is the fluorescence lifetime of endogenous fluorophores, which is an indicator of molecular microenvironment and structure changes, and features differences of biochemical and structural tissue composition between normal and neoplastic tissues. Lengthened decay state of the fluorescence helped increase photon counts and leaded to a greater SNR compared to standard time gating techniques. DOCI technique not only overcomes the challenges confounded by the local concentration of fluorophores, optical path, local excitation intensity or the local detection efficiency, which are common in autofluorescence imaging[21,23,36,37]. It also contrasts with previously explored FLIM techniques, which uses a confocal laser that results in long scanning time, small FOV, and limited depth of field that is impractical for intraoperative use.[16,26,27,38-40]

The DOCI technique has the potential to provide superior clinical outcomes such as decreased duration of operation, rates of recurrence, and associated costs compared to traditional frozen section analyses. Each HNSCC case averages about 5 frozen sections, which generally results in 15 minutes of processing time per margin.[3] Blind sampling that does not include residual tumors results in geographic misses. Further tissue processing and manipulation adds to the possible risk of technical error. Indeed, more than 80% of erroneous frozen section reads resulted from technical aspects of processing the tissue (e.g., superficial sectioning of buried tissue), rather than interpretation errors.[41] This may partly explain the 65% rate of persistent or recurrent disease in HNSCC patients despite a negative-margin resections.[42] Utilization of DOCI, therefore, would reduce operative duration, prevent sampling errors, and avert technical errors.

A significant portion of patients present with nodal metastasis of head and neck cancers but without an obvious primary site despite an extensive workup. This diagnosis of exclusion is known as head and neck squamous cell carcinoma of unknown primary (HNSCC UP) and is seen in about 5-10% patients with nodal metastasis. The incidence of HNSCC UP has been increasing; this is attributed largely to the rise in human papillomavirus-related cancers.[43] In these patients, blind biopsies of the presumed location of the primary tumor, namely the Waldeyer's ring surrounding the oropharynx, are performed. However, this can be costly and inefficient with multiple frozen section biopsies. Safe, non-invasive imaging techniques such as narrow band imaging have been proposed. However, these are largely operator dependent and rely on subjective interpretation of images by the surgeon.[3] Given the operator-independent nature of DOCI analysis, the DOCI technique is the superior alternate for rapid and reliable identification of primary site of tumor, or at least provides great assistance in the assessment of potential cancer.

The cancer prediction algorithm used with multiple channel information used in this study accurately detects tumor margins in an operator-independent manner. Previous studies utilized similar learning methods for cancer detection in the brain, lung and pancreas and many other medical applications[43-46] using various imaging modalities. In the DOCI technique, unique signals may be extracted by narrowing to specific bands of wavelengths. The use of combinations of band pass filters might facilitate unsupervised segmentation of nerve branches, apart from blood vessels and tumors. Combined with a basic anatomy database and a tracker, the interpreted DOCI images may serve as a guided-surgery system. This will not only aid in precision surgeries that improve the decision-making for margins involving critical nerves or blood vessels but also identification of nerve stumps that may be challenging to see in the operating field.

Although the DOCI technique elevates surgical precision to a new dimension, there are limitations to this study, including the lack of side-by-side comparison with traditional tissue biomarkers or previously published fluorophore-based imaging techniques. However, blinding the surgeon or the researcher for side-by-side comparison would be inherently not feasible given the requisite technical expertise to operate these systems at this time. Furthermore, the senior head and neck pathologist annotated the H&E stained slides with great confidence without the need for additional markers. With a more modularized and user-friendly interface, direct comparisons between intraoperative imaging techniques can be planned.

In summary, we developed and characterized a fast, wide field, label-free, multispectral imaging technique called DOCI, and we successfully validated the DOCI technique to detect cancer margins using head and neck SCC as model and histology as gold standard. The DOCI technique will be a useful first-line intraoperative diagnostic tool to guide surgical decision-making as an alternative to blind frozen section analyses or other operator-dependent imaging techniques. The use of this non-invasive imaging technique will enhance surgical outcomes and avoid the removal of uninvolved essential tissue to improve patients' quality of life. Future studies should include a larger scale research on various applications on this technology and embark on multi-site clinical trials.

## Methods

**Study design.** The goals of this study were to: (1) develop and characterize a fast, wide-field, label-free, multispectral imaging instrument for intraoperative autofluorescence lifetime imaging; (2) take image of the freshly resected patient tissues intraoperatively, compare the images with the corresponding H&E results, analyze the efficacy of the device in cancerous margin delineation and discuss the potential as a tool for biopsy guidance. We developed an imaging apparatus that balanced accuracy, resolution, imaging speed, field of view, working distance, depth of field and so on to overcome translational difficulties that traditional FLIM systems have. We then characterized the apparatus carefully using various dye solutions and a fluorescence spatial resolution target and demonstrated the feasibility of using the instrument to image freshly resected tissues from patients undergoing head and neck surgeries. Strict comparison with the gold standard revealed its potential to delineate cancerous margin and guide intraoperative biopsy sampling. This study was approved by the University of California, Los Angles (UCLA) Institutional Review Board.

**Dye solution preparation.** Three types of dyes at two concentrations were used to validate the DOCI system. The three dyes were Laurdan dissolved in dimethyl sulfoxide (DMSO), NADH in water, and 7-hydroxy-4-methylcoumarin in methanol. A drop of each solution was placed onto a microscope slide, and they were imaged under the same field of view by the DOCI system. Five solutions were prepared for fitting of data between DOCI values and absolute fluorophore lifetimes: 0.65 mM Laurdan dissolved in DMSO, 2.15 mM NADH in water, 0.4 mM 7-Hydroxy-4-methylcoumarin in methanol, 50 μM Prodan in methanol, and 50 μM Coumarin 2 in methanol. Each prepared solution was split into two equal parts for simultaneous fluorescence lifetime measurement using multiphoton microscopy and DOCI.

**Patient Selection.** This study was approved by the University of California, Los Angeles (UCLA) Institutional Review Board. 13 patients undergoing head and neck oncologic resection with curative intent were prospectively enrolled. Each patient provided written consent to enroll in the study before surgery. Specimens that contained only cancer or only benign tissues were excluded from the study, and those had been colored by ink were also removed from the data pool. Specimens from 7 patients were included in this study. Tumors were obtained from the following head and neck sites: ear (2), scalp (2), oral cavity (3). All specimens were confirmed to contain both squamous cell carcinoma and benign tissues on final pathology.

**Acknowledgement**

This study was supported by National Institutes of Health (R01CA205051, R01CA220663), American Academy of Otolaryngology-American Head & Neck Society Surgeon Scientist Career Development Award (MSJ) and the Tobacco-Related Disease Research Program of the University of California (MSJ). We thank the Translational Pathology Core Laboratory (TPCL) at UCLA and the Advanced Light Microscopy and Spectroscopy Laboratory at California NanoSystems Institute at UCLA for infrastructure and support and thank Ms. Chen Chen for illustrating the cartoon figures used in the paper.

**Author Contributions**

W.G., M.S.J. and O.S. conceived the original idea. Y.H. contributed to modeling, platform instrumentation, control software design, experimental validation, and data analysis. S.H. contributed to experimental validation and data analysis. S.M., A.Y.H, J.F.K. contributed to experimental validation. All the authors contributed to the clinical protocol design and to writing the manuscript.

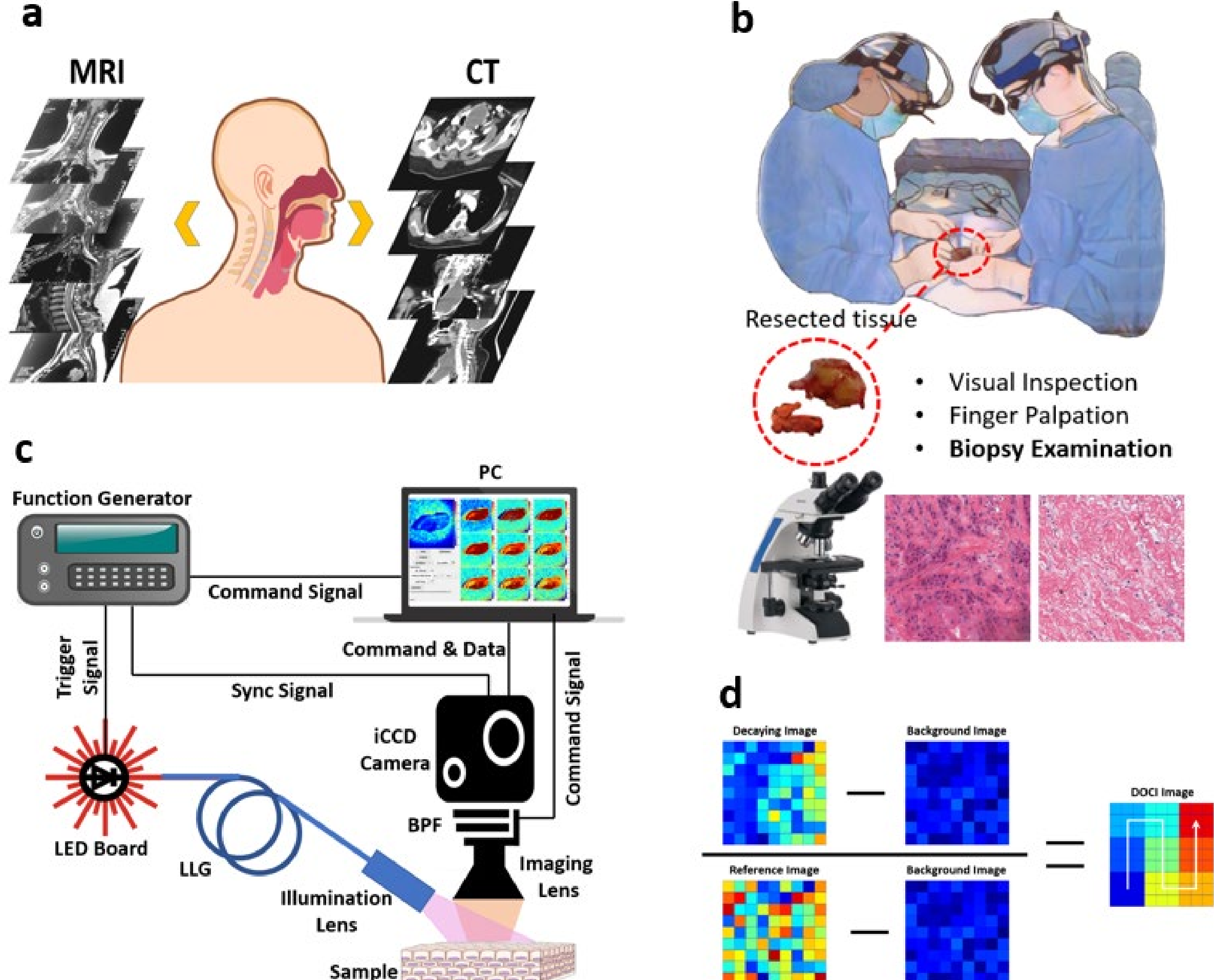

Figure 1. Description of the study background and working principle of the methodology. a, Patients undergoing head and neck cancer resection receive MRI, CT and other preoperative imaging examinations to assess tumor location, size and extent. b. Surgeons practice tumor resection based on preoperative imaging, visual inspection and tactile palpation. Final assessment of margins will be assisted with pathological examination of sampled biopsies. The resected tissue will be prepared for dynamic optical contrast imaging. c. High-level schematic of the dynamic optical contrast imaging instrument. LLG: light liquid cable; BPF: bandpass filter; iCCD: intensified change coupled device. d. Visualized imaging algorithm of dynamic optical contrast imaging.

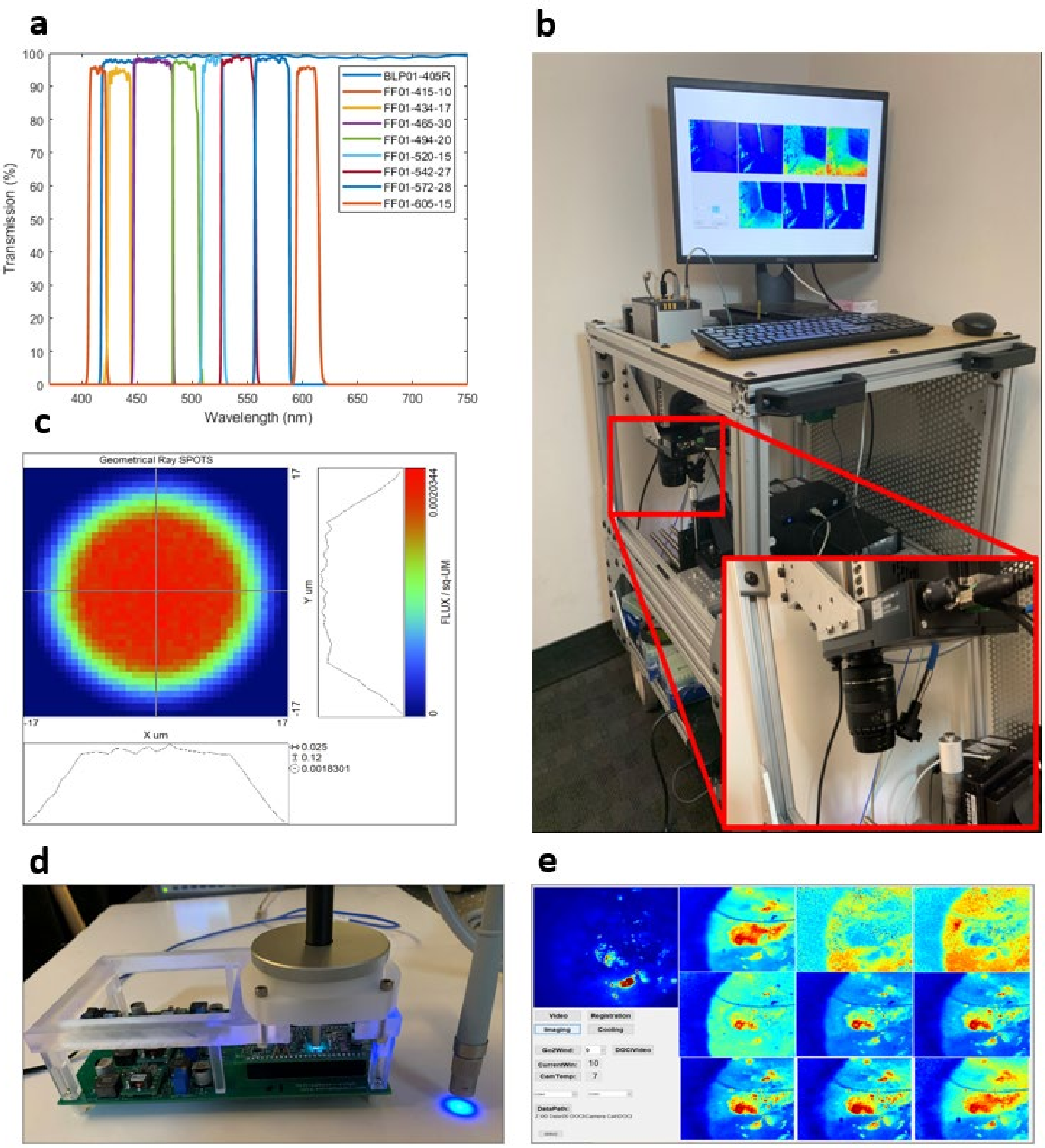

Figure 2. Instrumentation of the prototype system. a. Transmission spectrums of the filters applied in our current system. b. DOCI system used in this study with the imaging head apparatus zoomed in. c. Illumination intensity profile simulated with ASAP, showing a relatively homogeboues light filed out of theLLG d. Custom-made LED driver coupled to LLG drectly and outputs a circular illumination light field. e. The custom developed graphical user interface (GUI) displaying cancer tissue delineation, a thyroid cancer is imaged.

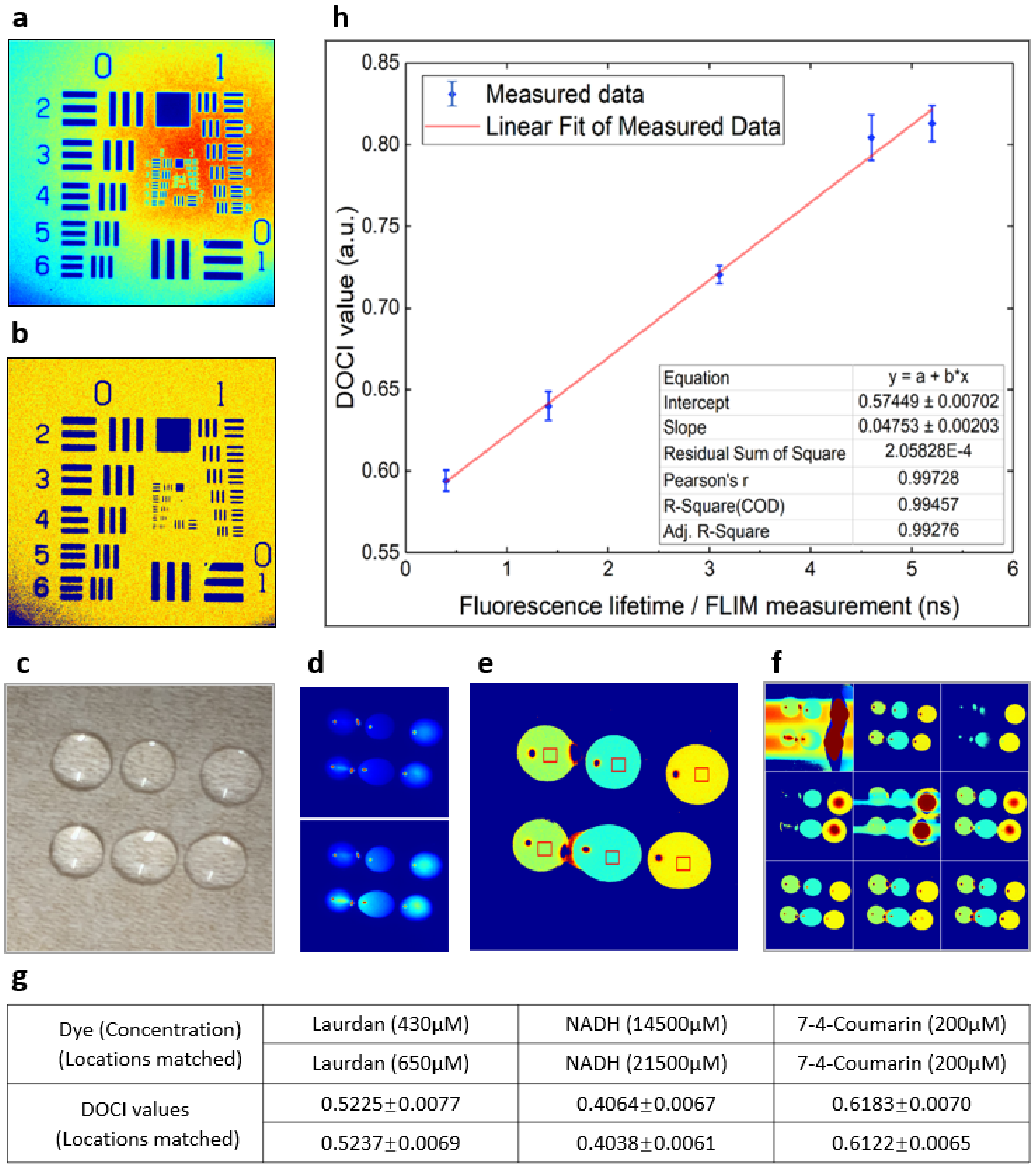

| Dye (Concentration) (Locations matched) | Laurdan (430μM) | NADH (14500μM) | 7-4-Coumarin (200μM) |
|---|---|---|---|
| | Laurdan (650μM) | NADH (21500μM) | 7-4-Coumarin (200μM) |
| DOCI values (Locations matched) | 0.5225±0.0077 | 0.4064±0.0067 | 0.6183±0.0070 |
| | 0.5237±0.0069 | 0.4038±0.0061 | 0.6122±0.0065 |

Figure 3. System characterization of the DOCI system. a, Fluorescence intensity image of USAF 1951 Resolution Targets. b, Fluorescence lifetime image of USAF 1951 Resolution Targets through filter 7 (528-556 nm). c, Brightfield image of dye drop samples on microscope slides. d, Fluorescence intensity images of dye drop samples in the decaying state (top) and reference state (bottom). e, DOCI image of dye drop samples through filter 8 (558-586 nm), with areas indicated for calculating the mean and standard deviation. f, DOCI images of dye drops from all nine filters (refer to the Methods section for wavelength details). g, Table showing dye types, concentrations, and DOCI values corresponding to their locations as shown in figure c. h, DOCI values compared to fluorescence lifetime values taken by a commercial FLIM system with five types of dye solutions.

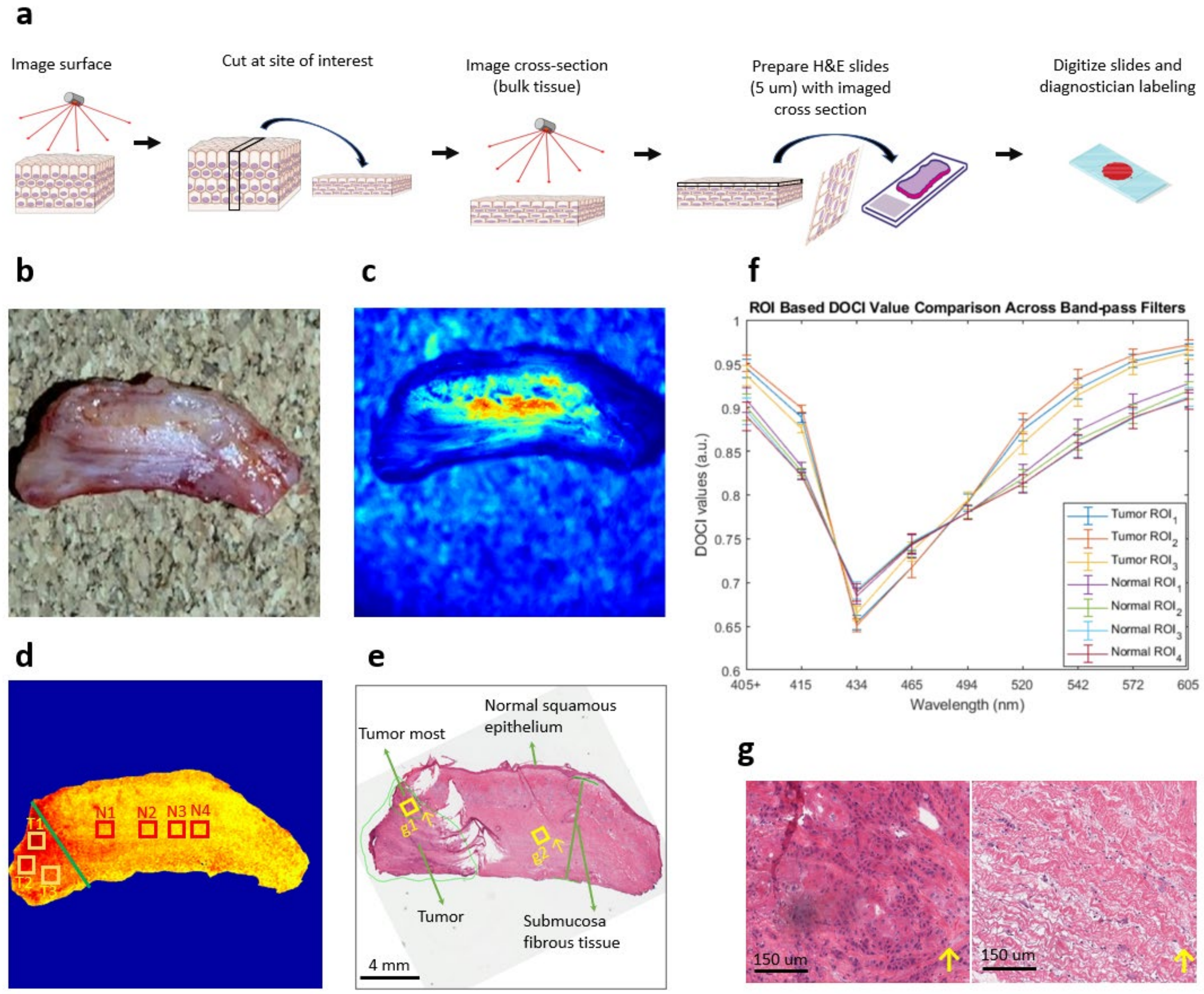

Figure 4. Intraoperative imaging of freshly sectioned diseased human tissue. a. Simplified protocol of tissue processing at TPCL and DOCI imaging. b. Cross-section image of the scalp tumor of a patient undergoing head and neck oncologic resection. c. Intensity image of the tissue with UV illumination. d. DOCI image of the tissue ($7^{th}$ window) with marked ROIs. e. Corresponding frozen section H&E slide with pathologist's annotation. f. ROI-based DOCI value comparison across studied wavelengths from 405-605 nm. g. High-power magnification of selected areas on the H&E slide as indicated g1 and g2 in figure e.

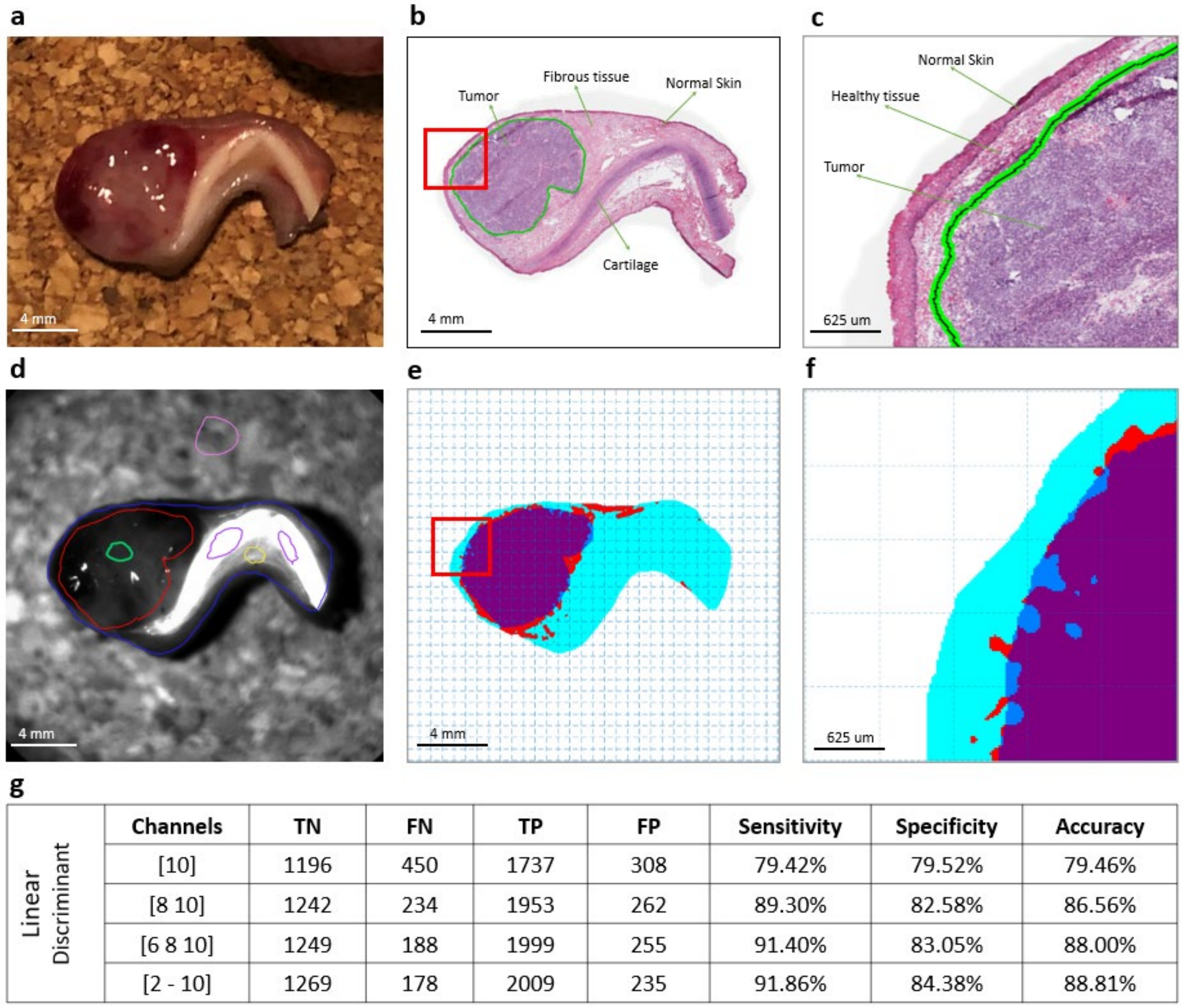

| Linear Discriminant | Channels | TN | FN | TP | FP | Sensitivity | Specificity | Accuracy |
|---|---|---|---|---|---|---|---|---|
| | [10] | 1196 | 450 | 1737 | 308 | 79.42% | 79.52% | 79.46% |
| | [8 10] | 1242 | 234 | 1953 | 262 | 89.30% | 82.58% | 86.56% |
| | [6 8 10] | 1249 | 188 | 1999 | 255 | 91.40% | 83.05% | 88.00% |
| | [2 - 10] | 1269 | 178 | 2009 | 235 | 91.86% | 84.38% | 88.81% |

Figure 5. Predicting cancerous area using multiple-channel lifetime information. a, A bright-field image of a grossly cut ear SCC. b, Corresponding H&E slides with cancerous area circled in green by the pathologist. c, High-power magnification of the boxed area in figure b, showing details of cancer boundaries. d, Registration of the DOCI image to the H&E image: blue for the area of the specimen, red for the ground truth of the cancers, green for randomly selected areas inside the red, purple for cartilage, yellow for fibrous, and pink for the corkboard background; e, Cancer predictions by trained classifiers with data from bandpass filters channels 2-9, with gridded blocks for efficacy evaluation. f, Zoomed-in scope of tissue predictions near the cancer margins. g, The summarized sensitivities, specificities, and accuracy of cancer tissue predictions made using the linear discrimination classifier based on samples from six patients.

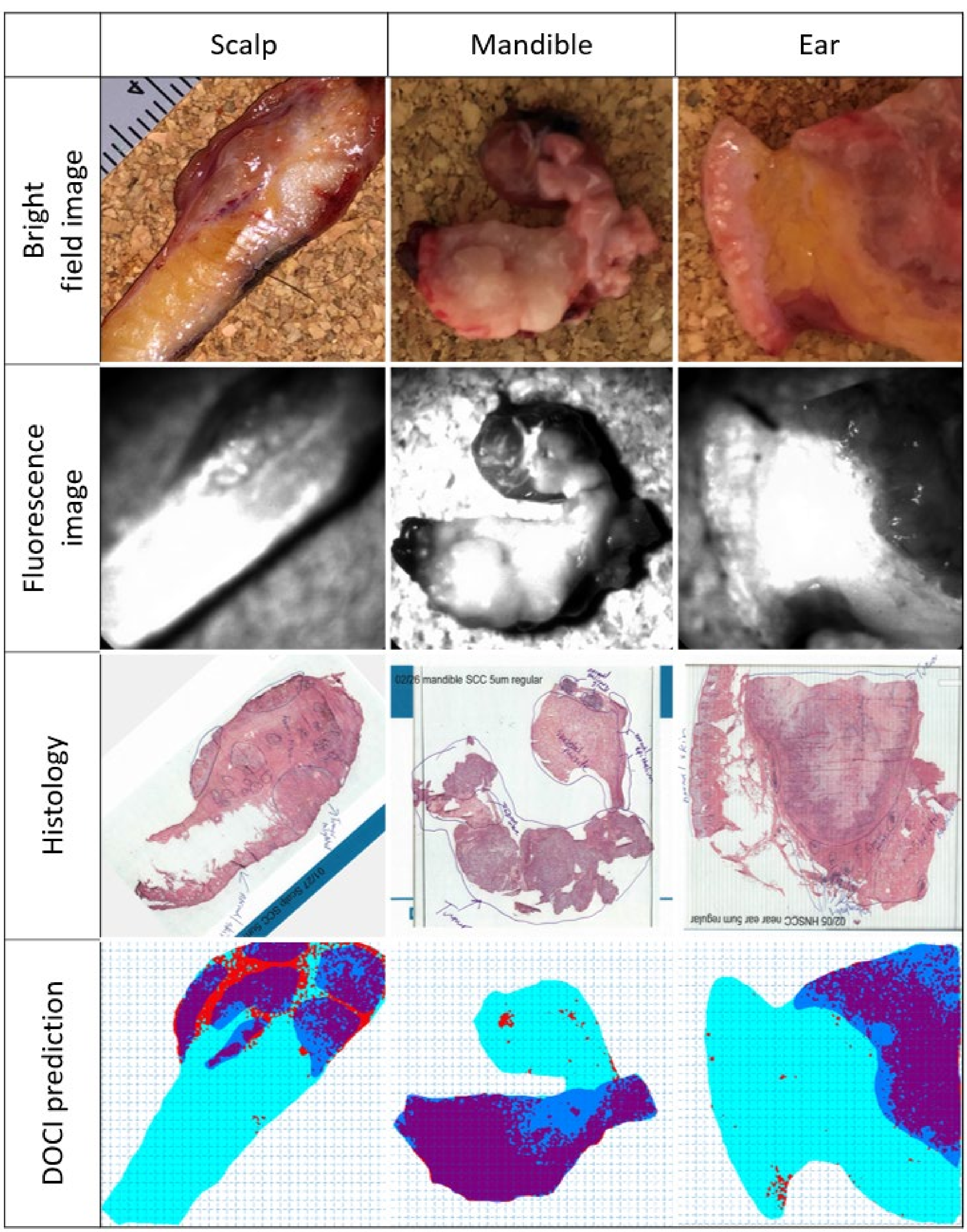

Extended Figure 1. Representative cases of cancerous tissues from different sites, scalp, oral and ear, with bright field images, fluorescence intensity images, corresponding histology annotations and DOCI prediction results displayed.

| | No. | Channels | TN | FN | TP | FP | Sensitivity | Specificity | Accuracy |
|---|---|---|---|---|---|---|---|---|---|
| Single channel results | 1 | [10] | 1196 | 450 | 1737 | 308 | 79.42% | 79.52% | 79.46% |
| | 2 | [9] | 1345 | 708 | 1479 | 159 | 67.63% | 89.43% | 76.51% |
| | 3 | [8] | 1334 | 699 | 1488 | 170 | 68.04% | 88.70% | 76.46% |
| | 4 | [7] | 1349 | 742 | 1445 | 155 | 66.07% | 89.69% | 75.70% |
| | 5 | [6] | 1310 | 772 | 1415 | 194 | 64.70% | 87.10% | 73.83% |
| | 6 | [2] | 1379 | 921 | 1266 | 125 | 57.89% | 91.69% | 71.66% |
| | 7 | [5] | 1260 | 908 | 1279 | 244 | 58.48% | 83.78% | 68.79% |
| | 8 | [3] | 1424 | 1186 | 1001 | 80 | 45.77% | 94.68% | 65.70% |
| | 9 | [4] | 1381 | 1148 | 1039 | 123 | 47.51% | 91.82% | 65.56% |
| Top 20 2-channel results | 1 | [8 10] | 1242 | 234 | 1953 | 262 | 89.30% | 82.58% | 86.56% |
| | 2 | [6 10] | 1198 | 203 | 1984 | 306 | 90.72% | 79.65% | 86.21% |
| | 3 | [7 10] | 1212 | 219 | 1968 | 292 | 89.99% | 80.59% | 86.16% |
| | 4 | [9 10] | 1236 | 249 | 1938 | 268 | 88.61% | 82.18% | 85.99% |
| | 5 | [5 10] | 1214 | 260 | 1927 | 290 | 88.11% | 80.72% | 85.10% |
| | 6 | [6 7] | 1244 | 323 | 1864 | 260 | 85.23% | 82.71% | 84.20% |
| | 7 | [2 10] | 1136 | 236 | 1951 | 368 | 89.21% | 75.53% | 83.64% |
| | 8 | [8 9] | 1309 | 447 | 1740 | 195 | 79.56% | 87.03% | 82.61% |
| | 9 | [6 8] | 1277 | 436 | 1751 | 227 | 80.06% | 84.91% | 82.04% |
| | 10 | [6 9] | 1301 | 497 | 1690 | 203 | 77.27% | 86.50% | 81.03% |
| | 11 | [3 10] | 1162 | 371 | 1816 | 342 | 83.04% | 77.26% | 80.68% |
| | 12 | [5 7] | 1252 | 462 | 1725 | 252 | 78.88% | 83.24% | 80.66% |
| | 13 | [4 10] | 1209 | 421 | 1766 | 295 | 80.75% | 80.39% | 80.60% |
| | 14 | [5 9] | 1311 | 524 | 1663 | 193 | 76.04% | 87.17% | 80.57% |
| | 15 | [7 9] | 1305 | 520 | 1667 | 199 | 76.22% | 86.77% | 80.52% |
| | 16 | [4 9] | 1330 | 549 | 1638 | 174 | 74.90% | 88.43% | 80.41% |
| | 17 | [2 9] | 1188 | 436 | 1751 | 316 | 80.06% | 78.99% | 79.63% |
| | 18 | [5 8] | 1284 | 543 | 1644 | 220 | 75.17% | 85.37% | 79.33% |
| | 19 | [4 8] | 1318 | 585 | 1602 | 186 | 73.25% | 87.63% | 79.11% |
| | 20 | [5 6] | 1237 | 523 | 1664 | 267 | 76.09% | 82.25% | 78.60% |
| Top 20 3-channel results | 1 | [6 8 10] | 1249 | 188 | 1999 | 255 | 91.40% | 83.05% | 88.00% |
| | 2 | [4 6 10] | 1208 | 151 | 2036 | 296 | 93.10% | 80.32% | 87.89% |
| | 3 | [6 9 10] | 1257 | 202 | 1985 | 247 | 90.76% | 83.58% | 87.84% |
| | 4 | [7 9 10] | 1250 | 197 | 1990 | 254 | 90.99% | 83.11% | 87.78% |
| | 5 | [3 7 10] | 1204 | 165 | 2022 | 300 | 92.46% | 80.05% | 87.40% |
| | 6 | [3 6 10] | 1191 | 154 | 2033 | 313 | 92.96% | 79.19% | 87.35% |
| | 7 | [4 5 10] | 1202 | 165 | 2022 | 302 | 92.46% | 79.92% | 87.35% |
| | 8 | [4 8 10] | 1230 | 196 | 1991 | 274 | 91.04% | 81.78% | 87.27% |
| | 9 | [4 7 10] | 1211 | 179 | 2008 | 293 | 91.82% | 80.52% | 87.21% |
| | 10 | [5 8 10] | 1245 | 218 | 1969 | 259 | 90.03% | 82.78% | 87.08% |
| | 11 | [5 9 10] | 1255 | 228 | 1959 | 249 | 89.57% | 83.44% | 87.08% |
| | 12 | [8 9 10] | 1243 | 216 | 1971 | 261 | 90.12% | 82.65% | 87.08% |
| | 13 | [5 6 10] | 1215 | 189 | 1998 | 289 | 91.36% | 80.78% | 87.05% |
| | 14 | [3 5 10] | 1206 | 183 | 2004 | 298 | 91.63% | 80.19% | 86.97% |
| | 15 | [3 8 10] | 1197 | 175 | 2012 | 307 | 92.00% | 79.59% | 86.94% |
| | 16 | [3 9 10] | 1214 | 193 | 1994 | 290 | 91.18% | 80.72% | 86.91% |
| | 17 | [4 9 10] | 1243 | 229 | 1958 | 261 | 89.53% | 82.65% | 86.72% |
| | 18 | [5 7 10] | 1248 | 234 | 1953 | 256 | 89.30% | 82.98% | 86.72% |
| | 19 | [7 8 10] | 1237 | 223 | 1964 | 267 | 89.80% | 82.25% | 86.72% |
| | 20 | [6 7 10] | 1218 | 216 | 1971 | 286 | 90.12% | 80.98% | 86.40% |
| All channels | 1 | [2 - 10] | 1269 | 178 | 2009 | 235 | 91.86% | 84.38% | 88.81% |

Extended Table 1. Efficacy statistics of DOCI predictions with different channel combinations, showing results of 9 one-channels, top 20 2-channels, top 20 3-channels, and all 9-channel.